\documentclass[a4paper]{jpconf}

\usepackage{prodint}  

 \usepackage{cite} 
\usepackage{amssymb} 
\usepackage{amsfonts} 
\usepackage{amsmath} 
\usepackage{slashed} 

\usepackage{egothic}%
\usepackage{pgothic}%
\usepackage{yfonts}%

 \usepackage{yhmath} 

\newcommand{\beq}{\begin{equation}}
\newcommand{\eeq}{\end{equation}}
\newcommand{\beqa}{\begin{eqnarray}}
\newcommand{\eeqa}{\end{eqnarray}}
\newcommand{\nn}{\nonumber}

\newcommand{\pbr}[2]{ \{ \hspace*{-2.6pt} [ #1 , #2\hspace*{1.4 pt} ] 
\hspace*{-2.6pt} \} }

\newcommand{\we}{\wedge}
\newcommand{\der}{\partial}

\newcommand{\inn}{\hspace*{2pt}\raisebox{-1pt}{\rule{6pt}{.3pt}\hspace*
{0pt}\rule{.3pt}{8pt}\hspace*{3pt}}}

\newcommand{\ka}{\varkappa}

\newcommand{\Psib}{\overline{\Psi}}
\newcommand{\Phib}{\overline{\Phi}}

\newcommand{\what}[1]{\widehat{#1}}

\newcommand{\bx}{{\mathbf{x}}}

\newcommand{\BPsi}{{\bf \Psi}} 
   
\newcommand{\BH}{{\bf H}}

\newcommand{\ugamma}{\underline{\gamma\!}}

\begin{document}
\title{\normalsize THE QUANTUM WAVES OF MINKOWSKI SPACETIME AND THE MINIMAL ACCELERATION FROM PRECANONICAL QUANTUM GRAVITY}

\author{I V Kanatchikov} 

\address{National Quantum Information Centre in Gdansk (KCIK), 
80-309 Gda\'nsk, Poland \\
IAS-Archimedes Project, C\^ote d'Azur, France
}


\begin{abstract}
We construct the simplest solutions of the previously obtained precanonical Schr\"odinger equation for quantum gravity, which correspond to the plane waves on the spin connection bundle and reproduce the Minkowski 
spacetime on average. Quantum fluctuations lead to the emergence of the minimal acceleration $a_0$ related to the range of the Yukawa modes in the fibers of the spin connection bundle. This minimal acceleration is proportional to the square root of the cosmological constant $\Lambda$ generated by the operator re-ordering in the precanonical Schr\"odinger equation. Thus the mysterious connection between the minimal acceleration 
$a_0$ in the dynamics of galaxies as described by Milgrom's MOND and the cosmological constant emerges as an elementary effect of precanonical quantum gravity. We also argue that the observable values of $a_0$ and $\Lambda$ can be obtained when the scale of the parameter $\varkappa$ introduced by precanonical quantization is subnuclear, in agreement with the previously established connection between the scale of $\varkappa$ and the mass gap in quantum SU(2) Yang-Mills theory. 
\end{abstract}

\section{Introduction}

The approaches to quantum gravity based on applying the standard methods of quantization to different versions of the gravitational Lagrangian \cite{kiefer,rovelli} almost inevitably lead to fundamental conceptual and technical problems of quantum gravity, such as the mathematical definition of the Wheeler-De Witt equation, the problem of time, the interpretation of (the measurement problem in) quantum cosmology, the problem of the correct classical limit in the loop quantum gravity, and the 122 orders of magnitude discrepancy between the theoretically plausible value of the cosmological constant and the observable one. 
 
 The approach called {\it precanonical quantization\/} \cite{ik2,ik3,ik4,ik5,ik5e} was proposed 
as a response to these  problems. It departs from the various forms of canonical quantization, which are based on the canonical Hamiltonian formalism with a necessarily distinguished time variable, and instead uses a 
 spacetime symmetric generalization of the Hamiltonian formalism from mechanics to field theory known in the calculus of variations under the name of the De Donder-Weyl (DW) Hamiltonian formulation  \cite{dedonder,weyl,kastrup}. 
The precanonical quantization is based on the Dirac quantization of the Heisenberg-like subalgebra of 
Poisson-Gerstenhaber brackets of differential forms representing dynamical variables, which were found in the DW Hamiltonian formulation in 
\cite{mybr1,mybr2,mybr3,mybr-pc,ik5} and further explored and generalized e.g. in \cite{helein1,helein3,paufler1,my-dkp}.


Quantization of brackets defined on differential forms naturally leads to  a hypercomplex generalization of quantum theory where operators and wave functions are Clifford-algebra-valued \cite{ik3,ik4,ik5,ik5e}.
The Clifford algebra in question is the complexified Clifford algebra of 
 spacetime. 
The DW Hamiltonian formulation and the quantization of Poisson-Gerstenhaber brackets of differential forms are 
spacetime symmetric by construction. 
No distinction between space and time variables is required. No notion of field configurations or their initial  or boundary data, i. e. the sections of the bundle whose base is 
spacetime and whose fibers are spaces where the fields take values, is required by the procedure of precanonical quantization. The quantum dynamics of fields is described using the sections of the Clifford bundle over the bundle of field variables $\phi^a$ over the 
spacetime with the coordinates $x^\mu$. These sections are called precanonical wave functions and, in general, have the form 
(in $n=1+3$ dimensions)
\beq 
\Psi (\phi^a,x^\mu)= \psi + \psi_\mu\gamma^\mu + \frac{1}{2!}\psi_{\mu\nu}\gamma^{\mu\nu}
+ ... + \frac{1}{4!}\psi_{\mu_1\mu_2...\mu_4}\gamma^{\mu_1\mu_2...\mu_4}
 . 
\eeq 
The field variables $\phi^a$ can be the Yang-Mills field variables $A_\mu^a$  \cite{ik5e,iky1,my-ymmg,iky3} 
or metric density variables $h^{\mu\nu}$ \cite{ikm1,ikm2,ikm3,ikm4},  
or tetrad variables $e^I_\mu$ \cite{iktp}, or spin connection variables $\omega_\mu^{IJ}$ \cite{ikv1,ikv2,ikv3,ikv4,ikv5}. 
The covariant analogue of the Schr\"odinger equation for the precanonical wave function has the form 
\cite{ikm1,ikm2,ikm3,ikm4,ikv1,ikv2,ikv3,ikv4,ikv5} 
\beq  \label{pseq}
i\hbar\varkappa \hat{\slashed \nabla} \Psi - \hat{H} \Psi =0 ,
\eeq
where $\hat{H}$ is the operator of the covariant analogue of the Hamiltonian in the DW Hamiltonian-like formulation: 
\beq
H:= \der_\mu {\phi^a} p^\mu_{\phi^a} -L, \quad  p^\mu_{\phi^a} := \frac{\der L}{\der \der_\mu \phi^a}, 
\eeq
 $\hat{\slashed \nabla}$ is the operator of the covariant Dirac operator on the 
  spacetime, 
 and the parameter $\varkappa$ is an ultraviolet quantity of the dimension of the inverse spatial volume. It appears on purely dimensional grounds given the fact that the physical dimension of classical $H$ is that of the mass density 
 (in  $c=1$ units used throughout the paper). 
Note that $\varkappa$ is also introduced in the course of the precanonical quantization when the Poisson-Gerstenhaber brackets 
are replaced by commutators, and the representation of operators corresponding to differential forms is constructed in terms of Clifford-algebra-valued operators. In particular, 
the 3-dimensional volume element $d\bx := dx^1\we dx^2\we dx^3$ is mapped to the Clifford algebra element 
\beq \label{qmap}
d\bx \mapsto \frac{1}{\ka}\ugamma{\, }_0 , 
\eeq
where $\ugamma{\, }_I$ denote the flat 
spacetime Dirac matrices, $\ugamma{\, }_I \ugamma{\, }_J + \ugamma{\, }_J\ugamma{\, }_I 
= 2 \eta_{IJ}$, 
$I,J = 0,1,2,3$.  This map is very similar to what is known as the ``quantization map'' or ``Chevalley map" in the Clifford algebra 
literature. 

The relation between the description of quantum fields in terms of Clifford-algebra-valued
precanonical wave functions $\Psi (\phi^a, x^\mu)$ and the standard QFT can be established if the latter is formulated 
in the Schr\"odinger functional picture \cite{hatfield}. In this picture, QFT is described in terms of time-dependent 
functionals of initial field configurations $\phi^a(\bx)$, $\BPsi([\phi^a(\bx)],t)$,  which obey the canonical 
Schr\"odinger equation 
\beq \label{sreq}
i\hbar \der_t \BPsi - \hat{\BH} \BPsi =0 ,  
\eeq
where $\hat{\BH} = \int\!d\bx \hat{T}{}^0_0$ is the operator of the canonical Hamiltonian.  

Taking into account that the precanonical wave function $\Psi (\phi^a,x^\mu)$ gives the probability amplitude of detecting the field value $\phi$ at the 
spacetime point $x$ and the Schr\"odinger wave functional $\BPsi([\phi^a(\bx)],t)$ 
is the probability amplitude of observing the field configuration $\phi(\bx)$ on the hypersurface of constant time $t$,  one can anticipate that the  Schr\"odinger wave functional is a continuous product, or product integral, over the spatial points $\bx$,  of precanonical wave functions restricted to the configuration $\Sigma$ given by $\phi = \phi(\bx)$, i.e. $\Psi_\Sigma (\phi = \phi(\bx), \bx, t)$,  and transformed from the representation with a diagonal $\hat{H}$ to the representation with a diagonal
 $\hat{T}^0_0 = \hat{H} - {\der_i \phi^a (\bx) \hat{p}{}^i_{\phi_a}}$.  The resulting 
 expression of the Schr\"odinger wave functional $\BPsi$ in terms of the precanonical wave 
functions $\Psi$ is given by the continuous product or the  Volterra product integral 
over $\bx$ denoted as $\prodi_\bx$: 
\beq \label{srprod}
\BPsi = {\sf Tr} \left \{\Prodi_\bx  
e^{-i \phi(\bx)\gamma^i\der_i \phi(\bx)/\ka} 
 \Psi_\Sigma ( \phi (\bx), \bx, t)
{}_{\mbox{ $\rvert$}\scriptscriptstyle
 \frac{1}{\ka}  
 \mapsto 
 \ugamma{\, }^0 d\bx } \right \} , 
\eeq
where the inverse of the quantization map (\ref{qmap}) is used as a natural step of transformation from the 
Clifford algebraic objects of precanonical quantization to the $\mathbb{C}$-valued functional of the canonical 
quantum field theory in the Schr\"odinger representation. In \cite{iks1,iks2} we have shown, 
for interacting scalar fields in Minkowski 
spacetime that $\BPsi$ constructed in equation (\ref{srprod}) satisfies the standard Schr\"odinger equation (\ref{sreq}) as a consequence of (the flat 
 spacetime version of) the precanonical Schr\"odinger equation (\ref{pseq}) restricted  to the surface of initial data $\phi^a = \phi^a(\bx)$ at the moment of time $t$.  A similar relation has been found also for quantum Yang-Mills theory on Minkowski spacetime \cite{iky1,iky3} and for scalar fields on curved spacetime \cite{iksc1,iksc2,iksc3}.  The existence of this relationship shows that standard QFT based on canonical quantization is the limiting case of QFT based on precanonical quantization corresponding to the inverse quantization map 
$\frac{1}{\ka} \ugamma{\, }_0 \mapsto   d\bx $ or, loosely speaking, to the limiting case of infinite $\ka$ corresponding to the unregularized volume of the momentum space, which equals $\delta^3(\mathbf{0})$. 


Let us also note that the existence of the spacetime symmetric  Hamilton-Jacobi (HJ) theory 
of fields which is associated with the DW Hamiltonian theory 
\cite{dedonder,weyl,kastrup,mohj1,mohj2,riahi} raises a question about the existence of a formulation of quantum field theories which reproduces the DW HJ theory in the classical limit. The precanonical quantization leads to such a formulation, at least in the case of scalar fields 
(cf. \cite{ik3,guiding}).

 In the previous papers, the precanonical quantization has been applied to general relativity in metric variables \cite{ikm1,ikm2,ikm3,ikm4}, 
 to the teleparallel equivalent of general relativity \cite{iktp}, and to general relativity in vielbein variables \cite{ikv1,ikv2,ikv3,ikv4,ikv5}. In this paper, 
 we will briefly outline the latter in Section 2 and then, in Section 3, construct 
 the solutions of the 
 precanonical Schr\"odinger equation for quantum gravity corresponding to the quantum version of Minkowski spacetime in Cartesian coordinates. 
 We will find that quantum effects lead to the emergence of minimal acceleration related to the range of the Yukawa modes of the precanonical wave function in the spin-connection space. We will show that this minimal acceleration is related to the square root of the cosmological constant.  
 The latter appears from the reordering of operators in the precanonical Schr\"odinger equation 
 for gravity. We will also obtain realistic estimations of both quantities, albeit with an error of several orders of magnitude, when the scale of the parameter  $\varkappa$ is below approximately 100 MeV, which is consistent with our previous rough estimation of the mass gap 
 in the quantum SU(2) gauge theory \cite{my-ymmg}.

\section{Precanonical quantum vielbein gravity }

In this section, we mainly collect together the key results from our previous considerations 
in \cite{ikv1,ikv2,ikv3,ikv4}. 
The construction of precanonical quantum vielbein gravity starts from the Einstein-Palatini Lagrangian density
\beq {\mathfrak L}= \frac{1}{8\pi G} {\mathfrak e} e^{[\alpha}_I e^{\beta ]}_J (\der_\alpha \omega_\beta^{IJ} +\omega_\alpha {}^{IK}\omega_{\beta K}{}^J) +\frac{1}{8\pi G}\Lambda {\mathfrak e} , 
\eeq
where the vielbein coefficients $e^{\alpha}_I$ and the spin connection coefficients $\omega_\alpha {}^{IK}$ are the independent field variables, and ${\mathfrak e} := \det(e^I_\mu)$. The DW Hamiltonian formulation leads to the constraints 
\beq
 {\mathfrak p}{}^\alpha_{\omega_\beta^{IJ}} :=\frac{\der {\mathfrak L} }{\der\, \der_\alpha{\omega_\beta^{IJ}}}\approx \frac{1}{8\pi G}
{\mathfrak e} e^{[\alpha}_Ie^{\beta ]}_{J }, 
\quad {\mathfrak p}{}^\alpha_{e^I_\beta} 
:=\frac{\der {\mathfrak L} }{\der\, \der_\alpha e^I_\beta}\approx 0 
\eeq
and the DW Hamiltonian density on the surface of constraints 
 \beq
 {\mathfrak H} := {\mathfrak p}{}_\omega\der \omega + {\mathfrak p}{}_e \der e - {\mathfrak L} 
 \approx -{\mathfrak p}{}^\alpha_{\omega_\beta^{IJ}}\omega_\alpha {}^{IK}\omega_{\beta K }{}^J 
 - \frac{1}{8\pi G}\Lambda {\mathfrak e}  . 
 \eeq
The constraints are second class according to the extension of the Dirac classification to the DW theory \cite{mydirac}. 
The calculation of the generalized Dirac brackets of forms representing the fundamental variables leads to the vanishing brackets 
of vielbeins and their polymomenta and very simple brackets of spin connection coefficients and their polymomenta, e.g.,  
\beqa \label{dbr11c}
{}&\pbr{{\mathfrak p}^\alpha_e}{e' \varpi_{\alpha'}}{\!}^D=0, \nn \\
{}&\pbr{p^\alpha_\omega}{\omega'\varpi_\beta}{\!}^D=
\pbr{p^\alpha_\omega}{\omega'\varpi_\beta} = \delta^\alpha_\beta \delta^\omega_{\omega'}, 
\label{dbr12c} 
\nn \\
{}&\!\!\hspace*{-10pt}\pbr{{\mathfrak p}^\alpha_e }{ {\mathfrak p}_\omega \varpi_{\alpha'}}{\!}^D 
\!=\pbr{{\mathfrak p}^\alpha_e }{\omega \varpi_{\alpha'}}{\!}^D 
\!=\pbr{{\mathfrak p}^\alpha_\omega}{e' \varpi_{\alpha'}}{\!}^D \!=0 , \,\label{dbr13c} \nn
\eeqa 
 where $\varpi_{\alpha} := \der_\alpha \inn dx^0\we dx^1\we ...\we dx^{3}$ is the basis of $3$-forms on $4$-dimensional spacetime. 
Quantization of these brackets according to the following generalization of Dirac's quantization rule 
\beq
[\hat{A}, \hat{B}] = - i\hbar \widehat{\mathfrak{e} \pbr{A}{B}{}^D}
\eeq 
 leads to the operator representations of the polymomenta of spin connection, the vielbeins, the 
DW Hamiltonian $H$, such that $\what{{\mathfrak H}}=: \what{{\mathfrak e} H}$, 
and the quantum Dirac operator:  
\begin{align}
\what{{\mathfrak p}}{}^{\alpha}_{\omega_\beta^{IJ}} &= -i\hbar\ka {\mathfrak e} \hat{\gamma}{}^{[ \alpha} 
\frac{\der}{\der \omega_{\beta]}^{IJ}} , \quad \mathrm{where} \quad  \hat{\gamma}{}^{ \alpha} := \hat{e}{}^\alpha_I\ugamma{}^I,  
  \label{paomop} \\
\hat{e}{}^\beta_I  & = -  8\pi i G \hbar\ka \ugamma{}^{J}\frac{\der}{\der \omega_{\beta}^{IJ}} ,\label{ebiop}
\\
 \what{H} &= 8\pi G \hbar^2\ka^2 \
 \ugamma^{IJ}  
\omega_\alpha{}^{KM}\omega_{\beta M}{}^L \frac{\der}{\der \omega_{\beta}^{KL}} \frac{\der}{\der \omega_{\alpha}^{IJ}}
- \frac{1}{8\pi G} \Lambda , \label{hdwoper}   \\
\what{\not\hspace*{-0.2em}\nabla} &= 
- 8\pi i G \hbar\ka \ugamma{}^{IJ}\frac{\der}{\der \omega_{\mu}^{IJ}} 
\left(\der_\mu +  \frac{1}{4}\, \omega_{\mu KL}\ugamma{}^{KL} \stackrel{\leftrightarrow}{\vee}\right) ,       
\label{nablaoper}
\end{align}
where $\stackrel{{\leftrightarrow}}{\vee}$   denotes the commutator 
(antisymmetric) Clifford product 
$
{\ugamma}{}^{IJ} \stackrel{\leftrightarrow}{\vee} \Psi 
 = \frac12 \left[\ \ugamma{}^{IJ}, \Psi\ \right] .
$
Hence the precanonical Schr\"odinger equation 
for quantum gravity takes the form 
\beq  \label{psegrav}
 {
\ugamma{\ }^{IJ}  \frac{\der}{\der \omega_{\mu}^{IJ}}   
   \left ( \der_\mu +   \frac{1}{4} \omega_{\mu KL}\ugamma{\ }^{KL} \stackrel{{\leftrightarrow}}{\vee}  
 - \frac{\der}{\der \omega_{\beta}^{KL}}
\omega_\mu{}^{KM}\omega_{\beta M}{}^L  \right)  
 \Psi (\omega,x)     
 +  \lambda \Psi (\omega,x) = 0 ,  
}
 \eeq
where 
$\lambda := \frac{\Lambda}{ (8\pi G \hbar\varkappa)^2}$ 
is a dimensionless combination of the fundamental constants of the theory, which depends on the operator ordering of 
$\omega$ and $\der_\omega$. 

Note that equation (\ref{psegrav}) was first obtained in \cite{ikv1} without  explicitly specifying  the 
action of the spin connection term on the wave function. The need for the commutator product 
was understood later in \cite{iksc1}, and the coefficient $\frac12$ in front of the commutator comes from the consideration of the Ehrenfest theorem similar to that in \cite{ik5e}, which is still unpublished.

The scalar product of precanonical wave functions is given by 
\beq \label{scprod}
\left\langle \Phi | \Psi \right\rangle 
:=  \Tr \int \Phib \, \what{[d\omega]}_{} \Psi , 
\eeq
where $\Psib:=\ugamma{}^0\Psi^\dagger\ugamma{}^0$    
and the operator-valued invariant integration measure on the 
  $24$-dimensional 
 space of spin connection coefficients 
\beq  \label{measop}
\what{[d\omega]}
 \sim  \hat{\mathfrak e}{}^{-6}\prod_{\mu IJ} d \omega_\mu^{IJ} .
\eeq 
The operator $\hat{\mathfrak e}{}^{-1}$ is constructed from the operators 
$\hat{e}{}^\beta_I$ in (\ref{ebiop}). 

Thus, we arrive at the ``spin connection foam" formulation of the geometry of quantum gravity in terms of the Clifford-algebra-valued wave function on the bundle of spin connection coefficients over spacetime, $\Psi(\omega, x)$,   and the transition amplitudes on the total space of this bundle, 
 $ \left< \omega, x | \omega', x'\right>$, which are the Green functions of (\ref{psegrav}). The 
 wave function corresponds to the 
 ``quantum fuzziness" of points of the total space and the 
  Green functions correspond to the quantum correlations between the points, 
 i.e. a quantum analogue of the classical connection. 

The normalizability of 
 precanonical wave functions: $\left\langle \Psi | \Psi \right\rangle < \infty$, 
leads to the vanishing contribution of the large curvatures 
 $R=d\omega + \omega\we\omega$   
to the probabilistic measure defined by the norm, 
 and that ensures the quantum-gravitational avoidance of a curvature  singularity by the precanonical wave function. 

In the context of quantum cosmology, $\Psi(\omega,x)$ defines the spatially homogeneous  
 statistics of local fluctuations of the spin-connection, which is classically given by the Hubble parameter $\dot{a}/a$,    not the ``distribution of quantum universes according to the Hubble parameter" as suggested by the picture of the superspace of 3-geometries emerging from the canonical 
 quantization. Hence the problem of the ``external observer" of the ``quantum ensemble of universes" disappears.  
 
The evolution of matter and radiation on the background of quantum gravitational fluctuations 
 whose statistics and correlations are predicted by (\ref{psegrav}) may 
  lead to predictable consequences for the distribution of 
 matter and radiation  at large cosmological scales, which may be observable. 
 
 In general, analysis of solutions of precanonical Schr\"odinger equation for quantum gravity, equation (\ref{psegrav}),  is a formidable task. In the following section, we will construct the simplest solutions which can be interpreted as a quantum wave counterpart  of the Minkowski spacetime.

\section{ Quantum wave states of Minkowski spacetime}

The Minkowski metric in Cartesian coordinates: 
\beq ds^2 = (dx^0)^2 - (dx^1)^2 - (dx^2)^2 - (dx^3)^2 , 
\nn \eeq
is characterized by the vanishing spin connection coefficients 
\beq \label{om}
 \omega_\mu^{IJ} = 0.
 \eeq
In this case, the precanonical Schr\"odinger equation (\ref{psegrav}) 
with $\Lambda = 0 $ takes the simple form 
\beq  \label{pseqm}
\ugamma^{IJ}\der_{\omega_\mu^{IJ}} \der_\mu \Psi = 0 .
\eeq

In terms of the plane waves on the total space $(x^\mu, \omega_\mu^{IJ})$ 
\beq
\Psi \sim e^{ik_\mu x^\mu + i \pi^\mu_{IJ} \omega_\mu^{IJ}} \tilde{\Psi} (k_\mu, \pi^\mu_{IJ}) 
\eeq
 we obtain 
 \beq \label{predis}
 \ugamma^{IJ} k_\mu \pi^\mu_{IJ} = 0 .
 \eeq
From (\ref{predis}), we obtain the dispersion relation 
 \beq  \label{dispers}
 k_\mu \pi^\mu_{IJ} k_\nu \pi^\nu{}^{IJ} = 0 , 
 \eeq  
which reflects a strong anisotropy due to the fibred  structure of the $(x,\omega)$ space. 

Therefore,  any solution of (\ref{pseqm}) has the form 
\beq\label{mi7}
 \Psi (\omega,x) = \int\! d^4k\! \int\! d^{24}\pi\ \delta(k_\mu \pi^\mu_{IJ} k_\nu \pi^\nu_{IJ})
e^{ik_\mu x^\mu + i \pi^\mu_{IJ} \omega_\mu^{IJ}} \tilde{\Psi} (k_\mu, \pi^\mu_{IJ}) .
\eeq

The solutions of interest should be normalizable. The normalizability on the 
subspace of vanishing spin connection coefficients $\omega = 0$ takes the form 
\beq \label{mi8}
\Tr \int d^{24}\omega \,  \delta^{24}(\omega) \Psib (\omega,x) \hat{\mathfrak{e}}{}^{-6} \Psi  (\omega,x) 
= \Tr (\Psib (0,x) \hat{\mathfrak{e}}{}^{-6} \Psi (0,x) ) = 1 ,  
\eeq
where the short-hand notations $\Psib (0,x)$ and $\hat{\mathfrak{e}}{}^{-6} \Psi (0,x)$ mean 
$\Psib (\omega,x)$ and $\hat{\mathfrak{e}}{}^{-6} \Psi (\omega,x)$ taken at $\omega = 0$. 

The states which lead to the Minkowski spacetime on the classical level have to satisfy the conditions 
\beq\label{mi9}
\langle \hat{g}^{\mu\nu}\rangle (x) = \Tr \int d^{24}\omega \delta^{24}(\omega) \Psib(\omega,x) 
\hat{\mathfrak{e}}{}^{-6}\hat{g}^{\mu\nu} \Psi (\omega,x) ) = \eta^{\mu\nu} , 
\nn 
\eeq
where the operator of the metric derived from the representation (\ref{ebiop}) has the form 
\beq \label{metrop}
\hat{g}^{\mu\nu} = - (8\pi G)^2 \hbar^2\ka^2 \eta^{IK} \eta^{JL} \der_{\omega_\mu^{IJ}} \der_{\omega_\nu^{KL}}.
\eeq  
Hence the wave functions which reproduce the Minkowski spacetime on average should satisfy 
\beq  \label{mi10}
\Tr ( \Psib(0,x)  \hat{\mathfrak{e}}{}^{-6} \hat{g}^{\mu\nu} ( \Psib(0,x) ) = \eta^{\mu\nu} . 
\eeq
In terms of the Fourier components, equation (\ref{mi10}) implies that 
\beq \label{mi11}
\tilde{\Psi} (\pi, k ) = \tilde{\Psi} ( -\pi, k ) .
 \eeq

By comparison with the normalizability condition, we conclude that 
\beq \label{mi12}
- (8\pi G)^2 \hbar^2\ka^2 \eta^{IK} \eta^{JL} \der_{\omega_\mu^{IJ}} \der_{\omega_\nu^{KL}} \Psi (0,x) = 
\eta^{\mu\nu} \Psi(0,x) .
\nn
\eeq 
Therefore, for the plane waves, 
\beq \label{mi13}
 (8\pi G)^2 \hbar^2\ka^2 \pi_{IJ}^\mu \pi^\nu{}^{IJ}{} = \eta^{\mu\nu} . 
\eeq 
Then, from the dispersion relation, it follows 
\beq\label{mi14}
\eta^{\mu\nu} k_\mu k_\nu = 0 . 
\eeq

Thus, the states corresponding to the (1+3)-dimensional Minkowski spacetime in the classical limit have: 
\begin{itemize}
\item the light-like modes $(\ref{mi14})$ along the spacetime dimensions (the base of the total space of the bundle of spin connection coefficients over spacetime); 

\item  4 massive (Yukawa) modes $(\ref{mi13})$ in the spin-connection spaces (the fibers of the total space of the bundle of spin connection coefficients over spacetime), 
which propagate in 6-dimensional subspaces with the 
coordinates $\omega_\mu^{IJ}$ for each $\mu = 0,1,2,3$; 

\item the range of those massive modes in $\omega$-space 
is $8\pi G \hbar \ka$, whose value we estimate below; 


\item the modes corresponding to the spatial $\mu = 1,2,3$ are tachyonic. 
Those  tachyonic modes, however, do not violate the causality in spacetime as they propagate along the fibers associated to each point of spacetime rather than in the spacetime itself. 
\end{itemize}

Note that the spin connection has the mass dimension +1, $\varkappa$ has the mass dimension +3, 
and the square of the Planck length $G\hbar$ has the mass dimension $-2$. Hence the range of the Yukawa modes in the spin connection space, $8\pi G \hbar \ka$, is given in the units of mass dimension +1, which is also the mass dimension of acceleration.

 If $\ka$ were Planckian, which is a seemingly natural first guess, then the range of Yukawa modes in the spin connection space is also Planckian, and they could be attributed to the quantum foaminess of spacetime at the Planck scale, as is usually assumed.  However, our study of quantum Yang-Mills theory from the perspective of precanonical quantization (see below) has produced evidence that $\ka$ is more likely a sub-nuclear scale quantity, which leads to a drastically different scale of the phenomena in question. 

\subsection{An estimation of $\varkappa$ from the mass gap in pure gauge theory }

From the Lagrangian of a pure non-abelian Yang-Mills theory we can derive the corresponding 
DW Hamiltonian function \cite{iky1,ik5e} and precanonically quantize it. It leads to the following 
expression for the DW Hamiltonian operator for the quantum pure YM field 
with the coupling constant $g$ \cite{iky1,ik5e,my-ymmg}
\beq \label{dwhop}
\what{H} = 
 \frac{1}{2} \hbar^2\varkappa^2 \frac{\der}{\der A_a^\mu\der A^a_\mu } 
- \frac{1}{2}ig\hbar\varkappa  C^a{}_{bc}A^b_\mu A^c_\nu 
\gamma^\nu \frac{\der}{\der A^a_\mu } \; . 
\eeq
The fact that the eigenvalues of the DW Hamiltonian operator for the pure YM field yield the spectrum of masses of the  propagating modes is manifested in the precanonical Schr\"odinger equation in flat spacetime
\beq \label{nsequ}
i\hbar \gamma^\mu\der_\mu \Psi = \frac{1}{ \ka } \what{H}\Psi . 
\eeq
In \cite{iky3}, we have shown that equations (\ref{dwhop}) and  (\ref{nsequ}) for the wave function 
$\Psi (A,x)$ reproduce the functional Schr\"odinger equation for the wave functional 
$\BPsi ([A(\bx)],t)$ after the (3+1) decomposition and the ``dequantization map" 
$\frac{1}{\ka}\ugamma_0 \mapsto \varpi_0 = d\bx$ (cf. (\ref{qmap})). 

In the temporal gauge ${A}{}^a_0 = 0$, we can limit ourselves to the operator (\ref{dwhop})  
written only in terms of the spatial components $A^a_i$. For SU(2) theory with $a,b,c=1,2,3$ and $i,j=1,2,3$, we were able to estimate the gap between the ground state and the first excitation 
with the vanishing non-abelian charge (a ``color" or rather ``isospin" in the context of SU(2))
\beq
\langle \frac{1}{\ka}\hat{H} \rangle > \left( \frac{8g^2\hbar^4\ka}{32}\right)^{1/3} |\mathsf{ai}'_1|  , 
\eeq
where $\mathsf{ai}'_1$ is the first root of the derivative of the Airy function. This gap in the spectrum of the DW Hamiltonian operator can be identified with the mass gap 
\beq \label{mgap}
\Delta \mu \approx 0.86 (g^2\hbar^4\ka)^{1/3} .
\eeq
Therefore, the scale of $\ka$ is close to the scale of the mass gap.
For SU(3) YM theory, this formula  will have a different numerical coefficient in front and a different value of the coupling constant $g$. The SU(3) QCD mass gap lies between the pion masses at 130 MeV and the alleged glueball masses at a few GeV. The numerical factor  in (\ref{mgap}) for SU(3) may change  several times, and the coupling constant $g$ is of the order $10^{-2} - 10^{-1}$ (in the units of $\sqrt{\hbar}$).  With those uncertainties, we estimate $\ka^{1/3}$ is below 1 GeV with an  error of up to 2 orders of magnitude.

\subsection{The minimal acceleration} 

With the GeV-scale $\ka$ we obtain 
\beq\label{minacc}
8\pi G \hbar\ka \sim   10^{-23 \pm 3\times 2}\ \mathrm{cm}^{-1} . 
\eeq
This quantity is compatible with the scale of the Hubble radius $R_H \sim  10^{28}\ \mathrm{cm}$ and 
the cosmological constant $|\Lambda | \sim 10^{-56}\ \mathrm{cm}^{-2}$,  if the scale of $\ka$ is below 100 MeV, which is on the edge of our margin of error. In this case, $8\pi G \hbar\ka$ coincides with the scale of the minimal acceleration $a_0 \sim \sqrt{\Lambda}$ which is known from Milgrom's theory of MOND \cite{mond1,mond2,mond3,mond4,mond5}. Note that the Yukawa modes in the spin-connection space, which set the threshold of acceleration $8\pi G \hbar\ka$, emerge from quantum fluctuations of spin connection around the vanishing value of the spin connection of Minkowski spacetime. They establish the limit below which quantum fluctuations of spacetime violate the notion of acceleration-less inertial frames which underlies classical Minkowski spacetime. Note also that our value of the minimal acceleration  appears here in the context of a quantum analogue of the Minkowski spacetime and it  may slightly change for the quantum analogues of more realistic cosmological spacetimes.

\subsection{The cosmological constant} 

We have already pointed out that the expression of the 
 precanonical Schr\"odinger equation for gravity (\ref{psegrav})  is defined up to the operator ordering of 
$\omega$ and $\der_\omega$. A reordering in the spin connection term 
will produce a constant of the order $\frac{1}{4}4^3$ added to the dimensionless 
$\lambda = \frac{\Lambda}{(8\pi G\hbar\ka)^2}$ constructed from the bare cosmological constant $\Lambda$. If the latter equals zero, then the contribution to the cosmological constant from the reordering of operators, i.e. essentially from the quantum fluctuations of the spin connection, can be estimated as $\Lambda_\omega \sim 4^2 (8\pi G\hbar\ka)^2$.  For $\ka \sim 10^{0\pm 3\times 2}$ GeV$^3$,  in agreement with the estimation in Section 3.1, we 
obtain $\Lambda_\omega \sim 10^{-45 \pm 2\times 6}$ cm$^{-2}$. This estimation is again consistent with the observed value of the cosmological constant  if the scale of $\ka$ is below approximately 100 MeV. In this case, the minimal acceleration in (\ref{minacc}) is related to $\Lambda_\omega$ as follows 
\beq
a_0 \approx \frac{1}{4} \sqrt{\Lambda_\omega}, 
\eeq
which is close to the current observed value $a_0 \approx 1.2 \times 10^{-10}$ m$\cdot$s$^{-2}$ or $a_0\approx 10^{-29}$ cm$^{-1}$ in the 
$c=1$ units. Thus the mysterious connection between the phenomenological minimal acceleration in the dynamics of galaxies as described by MOND (as an alternative to the dark matter) and the cosmological constant (as the simplest dark energy) emerges 
as an elementary effect of precanonical quantum gravity. 

\ack  I gratefully acknowledge V.A. Kholodnyi for his interest and insightful comments, 
 J. Kouneiher for his interest and encouraging support, and M. Wright for his remarks helping to improve the linguistic quality of the manuscript. I also thank Hans-Thomas Elze for his 
tireless organization of DICE Conferences in Castiglioncello, which provided an inspirational framework for exchanging ideas.

\end{document}